\documentclass[groupaddress,prl,nofootinbib,preprintnumbers]{revtex4}

\usepackage{amsfonts}
\usepackage{mathrsfs}
\usepackage{amssymb}
\usepackage{amsmath}
\usepackage{graphicx,epsfig}

\def\lsim{\mathrel{\mathpalette\@versim<}}

\def\gsim{\mathrel{\mathpalette\@versim>}}

\begin{document}

\title{Probing Lorentz Violating (Stringy) Quantum Space-Time Foam }

\author{Nick E. Mavromatos}
 \affiliation{King's College London, Department of Physics, London WC2R 2LS, U.K.}

\begin{abstract}
 Quantum Space Time may be characterized by a plethora of novel phenomena,
 such as Lorentz violations and non-trivial refractive indices, stochastic metric fluctuation effects leading to decoherence of quantum matter and non-commutativity of space-time coordinates. In string theory, which is one of the major approaches to quantum gravity, such coordinate non-commutativities arise naturally in many instances. In the talk I review one such instance, which arises in the modern context of D-brane defects in the background space time, over which string matter propagates. This serves as a prototype of space-time foam in this context. I chose this model, over many others, because it may actually have some unique features that can be falsified experimentally either by means of high-energy astrophysical observations or in some particle-interferometers, such as neutral meson factories. In particular, the model may explain the recent observations of the FERMI Gamma-Ray Telescope on delayed emission of 30 GeV photons from a distant Gamma-Ray-Burst 090510, in agreement with previous observations from the MAGIC and HESS Telescopes, but can also lead to falsifiable predictions for quantum foam effects in forthcoming upgrades of certain ``particle interferometers'', such as neutral meson factories.

\end{abstract}

\maketitle


\section{Introduction}

The theory of quantum gravity, that is a consistent quantum theory explaining the microscopic structure of space-time itself and its dynamics,  is still largely unknown. Although there are several major theoretical approaches to the subject, aiming either at a fundamental understanding of its mathematical foundations, such as canonical quantization, loop quantum gravity, spin foam and string theory, or at a more phenomenological level, such as
Deformed Special Relativities or effective theories in stochastic space-times, entailing modified dispersion relations for matter probes, decoherence and CPT Violation, nevertheless the distinct lack of experimental evidence for quantum gravity so far makes the task of selecting the physically relevant model(s) a formidable one.

From a fundamental physics viewpoint, understanding a theoretical model of quantum gravity would necessitate knowledge of the fundamental space-time symmetries that characterize it. The possible breaking of such symmetries would in principle lead to observable effects. One of the most fundamental symmetries of space-time is that of Lorentz. This is the cornerstone of Special Relativity on which the modern particle physics field theory and phenomenology are based.
However, Lorentz Invariance may not be a true symmetry of quantum gravity, that is one may have tiny violations of the symmetry at the characteristic length scale of the quantum-gravitational interactions, which in the traditional
four-dimensional approach is believed to be close to the Planck length $\ell_P = 10^{-35}$~m. However, in the modern context of higher-dimensional theories of gravity, such as string theory, this scale may be as low as the length corresponding to an energy of a few TeV.

A concrete, but by no means generic, framework for studying issues of Lorentz symmetry violation is provided by models of quantum gravity which entail a ``foamy'' structure of space-time, the latter being microscopic black holes, or other topologically non-trivial space-time configurations, and in general space-time defects.
The presence of a defect can break Poincar\'e and/or Lorentz invariance of the vacuum of quantum gravity, which in this instance resembles a ``medium'' over which ordinary matter propagates. In such ``media'', which are stochastically fluctuating, the photon propagation is characterized by non-trivial vacuum refractive indices~\cite{aemn,nature,mestres}. This constitutes, therefore, a concrete manifestation of Lorentz symmetry violation.
There is a plethora of theoretical models entailing such violations, most of which can be studied by the method of effective Lagrangians~\cite{myers}. Specifically, one writes down higher-dimensional and higher-derivative local non-renormalizable operators, violating Lorentz and/or CPT symmetries, in a flat space-time set up, and then (s)he studies various effects owing their existence to potential Lorentz Violation in those phenomenologically motivated field theories. In this way, the coefficients of these higher-dimensional terms are constrained by observations. The strength of these Lorentz violations depends crucially on the power of the Quantum-Gravity mass scale, $M_{\rm QG}$, which the anomalous effects in the matter dispersion relations are suppressed by. There are models entailing \emph{linear} suppression, quadratic, \emph{etc.} The linear effects seem to have been ruled out already, if they are universal among particle species. This has been made possible by means of astrophysical observations of the synchrotron radiation~\cite{crab}. Indeed, if the electron dispersion relations are modified by terms $E^2/M_{\rm QG}$, where $E$ is the energy of the electron probe, then observations on the position of the maximum of the synchrotron radiation spectrum in distant nebulae, due to the existing magnetic fields, imply severe restrictions on $M_{\rm QG}$, requiring it to be larger than the (four-dimensional) Planck scale $M_P \sim 10^{19}$~GeV by several orders of magnitude. In fact, the sensitivity of the observations reaches the Planck scale for quadratic modifications to the dispersion relations, $E^3/M_{\rm QG}^2$.

However, quantum gravity may not act universally among particle species, especially as far as the quantum-foam aspects are concerned. This sort of ``violation'' of equivalence principle has been realized~\cite{ems} in a concrete model of space-time foam in the modern context of string/brane theory~\cite{Dfoam}, and it will be the topic of our discussion below. The reason why I chose this example is twofold: firstly, because it exhibits some quite unusual features, which evade the above-mentioned stringent astrophysical constraints of linear models, since the action of the medium of quantum gravity on the electrons and photons is quite distinct. Secondly, as we shall discuss here, the model makes concrete predictions
on the modification of the correlations of entangled states in some particle interferometric devices, such as neutral meson factories, which in fact may be uniquely associated only with this type of quantum gravity models.

\section{Probing Quantum-Gravity Medium effects with Astrophysical Observations}

The idea is simple~\cite{aemn,nature,mestres}: if the quantum space-time vacuum behaves as a stochastic medium, then modified dispersion relations for matter probes may arise, in a similar spirit to the non-trivial refractive indices of photons in an ordinary medium.
However, unlike normal matter, a quantum space time foam leads to anomalous dispersion effects that increase proportional to some power of the energy of the probe (in units of the speed of light in vacuo of Special Relativity $c=1$):
\begin{equation}\label{mdr}
E^2 = p^2 \left(1 + \sum_{n=1}^\infty a_n \left(\frac{p}{M_{\rm P}}\right)^n \right) + m^2
\end{equation}
where the coefficients $a_n$ are algebraic and depend on the details of the microscopic model, and $M_P \sim 10^{19}~{\rm GeV}$ is the Planck mass, thought of as a typical scale at which quantum-gravity effects set in.
For all practical purposes, at least so far, particle probes have energies much lower than $M_P$, and in this sense, only the leading non-trivial order in the series (\ref{mdr}) plays a r\^ole relevant for phenomenology, and thus the fitting formulae include either the $n=1$ or $n=2$ terms only in all studied cases so far.

A modified dispersion relation (\ref{mdr}) for photons ($m=0$) implies a non-trivial refractive index, for an anomalous-dispersion effect of order $n=1$ or $2$, with photon energies $E \ll M_P$ :
\begin{equation}
\eta  = 1 - \frac{1}{2} a_n \left(\frac{E}{M_P}\right)^n~,
\label{refrindex}
\end{equation}
where, to leading order, $p \simeq E$.
Negative coefficients $a_n \, < \,  0$ imply \emph{sub-luminal} refractive indices.

Several models of quantum gravity~\cite{amelino}, do predict such modified dispersion relations (MDR). Historically, the first concrete model appeared in the context of string theory, and in particular of non-critical strings~\cite{aemn}, where the anomalous (notably, sub-luminal) dispersion effects for photons have been associated with deviations from world-sheet conformal invariance. The latter is in turn linked to non-trivial interactions of stringy matter with space-time defects. Models of Loop Quantum Gravity~\cite{lqg} and Deformed Special Relativities (DSR)~\cite{dsr,smolin} are not necessarily characterised by such MDR, although some predictions on MDR have been made in certain models fairly early on, after the initial suggestion of \cite{aemn,nature}.
We should note that the models of loop quantum gravity predicting MDR~\cite{gambini} are based on the assumption of the existence of a special vacuum state, which, however, has been questioned. Moreover in DSR models there are situations in which the physically measurable momenta do not necessarily exhibit any MDR.

Nevertheless, the phenomenology of such MDR, which I must stress once again is \emph{only one aspect} of possible experimental signatures of quantum, gravity, is very rich, ranging at present from observations of the arrival times of photons from distant astrophysical sources, such as Active Galactic Nuclei (AGN)~\cite{MAGIC,hess,MAGIC2,hessnew} and Gamma-Ray-Bursts (GRB)~\cite{fermi}, to synchrotron radiation spectral studies from distant nebulae~\cite{crab}, as well as studies focused on the absence of
birefringence effects of astrophysical photons~\cite{uv,grb} and ultra-high-energy photons, with energies higher than $10^{20}$ eV, and related phenomena~\cite{sigl}.

To understand the basic idea behind these phenomenological studies, we mention that, for photons of different energies, emitted ``simultaneously'' (i.e. within the experimental accuracy of measured delays) from an astrophysical source, quantum-gravity-induced sub-luminal refractive indices would lead to delayed arrival of the more energetic photons~\cite{nature}. The expansion of the Universe affects the ``measured delay'', $\Delta t_{\rm total}^{\rm obs}$, at the observation point~\cite{naturelater,JP} as follows:
\begin{equation}
  \Delta t_{\rm total}^{\rm obs} = H_0^{-1}\, \int_0^z \frac{a_n}{2}\left(\frac{E_{\rm obs}}{M_P}\right)^n (1 + z')^n \frac{d z'}{\sqrt{(1 + z')^3 \Omega_M + \Omega_\Lambda}}~.
\label{totaldelayobs}
\end{equation}
where $z$ is the red-shift, ${H}_0 \sim 2.5 \times 10^{-18} ~{\rm s}^{-1}$ is the present epoch Hubble expansion rate
and $\Omega_{M(\Lambda)}$ is the present-era energy density of matter (vacuum), in units of the critical density of the universe. Above, we have assumed the standard cosmological model, according to which there is 29\% Matter, including Dark Matter, $\Omega_M =0.29$, and 71\% Dark Energy, in the form of a cosmological constant, $\Omega_\Lambda = 0.71$. In the event that the foam effects are the dominant cause of the time delays, one can then obtain information on the magnitude of the quantum-gravity ``effective'' scale,
\begin{equation}
M_{{\rm eff}, QG}^n = \frac{2\,M_P^n}{|a_n|}~, \quad n=1 ~{\rm or}~ 2~.
\label{effqgscale2}
\end{equation}
On the theoretical front, some authors attempted to use the \emph{formalism of local effective field theories} (LEFT) in order to describe the quantum-gravity-induced refractive indices. According to such a formalism~\cite{myers}, one can write down higher-derivative, non-renormalizable, local operators, which are added to the standard model lagrangian, in a Minkowski space-time background, and which are Lorentz violating, leading to modified dispersion relations (in a phenomenological manner) of the form (\ref{mdr}).
In this context, the calculated delays are dependent on the polarization of the photon, and one has both super-luminal and sub-luminal effects~\cite{myers}, leading unavoidably to \emph{birefringence}, that is, refractive indices depending on the photon polarization.

Such studies have imposed pretty stringent limits on the coefficients of the linearly suppressed (with $M_{\rm QG}$) anomalous dispersion effects for several probes, in particular electrons, from studies of the synchrotron radiation of Crab Nebula~\cite{crab}. Specifically, for electrons one has that $a_1^{\rm electron} \le 10^{-17}$, while for  photons, the absence of birefringence effects from a plethora of measuremtns~\cite{uv,grb,crab} implies $a_1^{\rm photon} \le 10^{-8}$.  In fact, for electrons, the sensitivity of the synchrotron radiation studies~\cite{ems} is close to Planck scale $M_P \sim 10^{19}$~GeV, for quadratically suppressed MDR, $n=2$. Similar stringent bounds on sub-luminal refractive indices for photons and MDR for electrons may be obtained from the fact that ultrahigh energy photons, with energies higher than $10^{20}$ eV have not yet been observed~\cite{sigl}, whose existence is allowed in models with Lorentz Violating MDR effects.

One should add to these effects the very recent observation~\cite{grb090510} from the FERMI Telescope
of an extremely short burst, GRB 090510, of duration of order of a second, with highest-energy  photons with energies $31$ GeV, arriving at about $0.4$ sec later than lower energy photons, of energies of order 1 GeV or in the MeV range. Assuming that the dominant reason for the high energy photon delay is a sub-luminal refractive index of quantum gravity type, one arrives~\cite{grb090510} at the result that, for linear Planck suppression, the coefficient $a_1^{\rm photon} \ll 1$, thereby exceeding Planck scale sensitivity.

Many, interpreted such results as implying absence of linearly suppressed MDR, and hence the exclusion of all theoretical models that predict them, since on naturalness grounds one would expect the coefficients $a_1$ to be of order one.
However, things are not so simple, and in fact the LEFT formalism may not represent accurately a true stochastic quantum gravity situation. As I will discuss in the next section, there are models for stringy quantum space-time ``foams'', which go beyond LEFT, thereby avoiding  the above-mentioned stringent constraints coming from the non-observation of birefringence. Moreover, in such models: (i) quantum gravity effects are not universal among particle probes, thus avoiding the stringent constraints from synchrotron radiation studies from Crab Nebula~\cite{crab} and (ii) the models predict linearly suppressed MDR, with sub-luminal refractive indices for photons, but such that the coefficients $a_1^{\rm photons}$ can be several orders of magnitude smaller than unity in a quite natural set up, in agreement with the recent FERMI observations on GRB 090510~\cite{grb090510}. Such models can be found within the modern context of string theory, involving D-brane defects in the (higher-dimensional) space time. I will also discuss in this talk some other, rather unique, predictions of these models concerning the (modified) quantum correlators of entangled states in particle interferometers.

\section{Brane Models of Space-Time Foam}

\begin{figure}[ht]
\centering
\includegraphics[width=7.5cm]{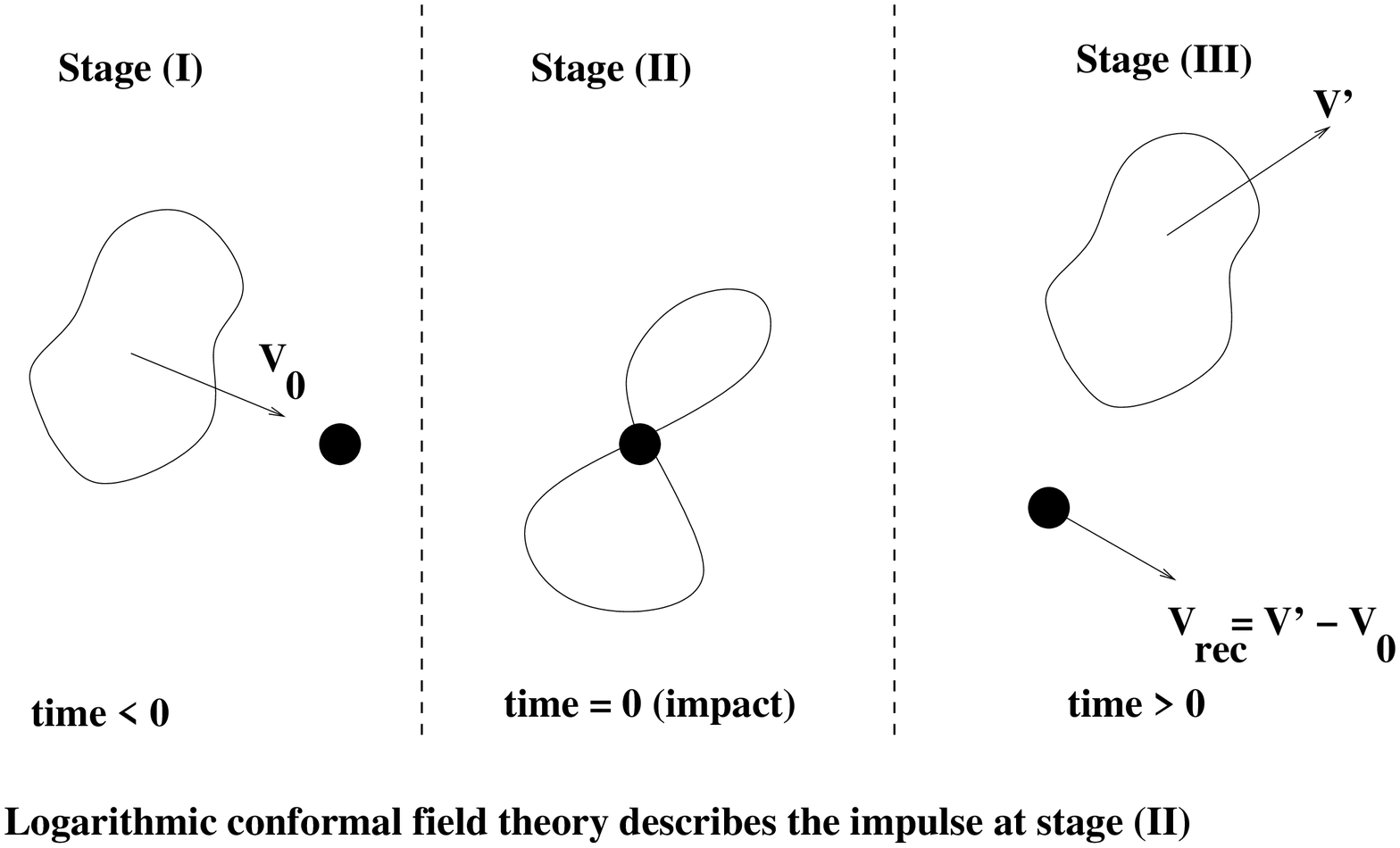} \hfill
\includegraphics[width=7.5cm]{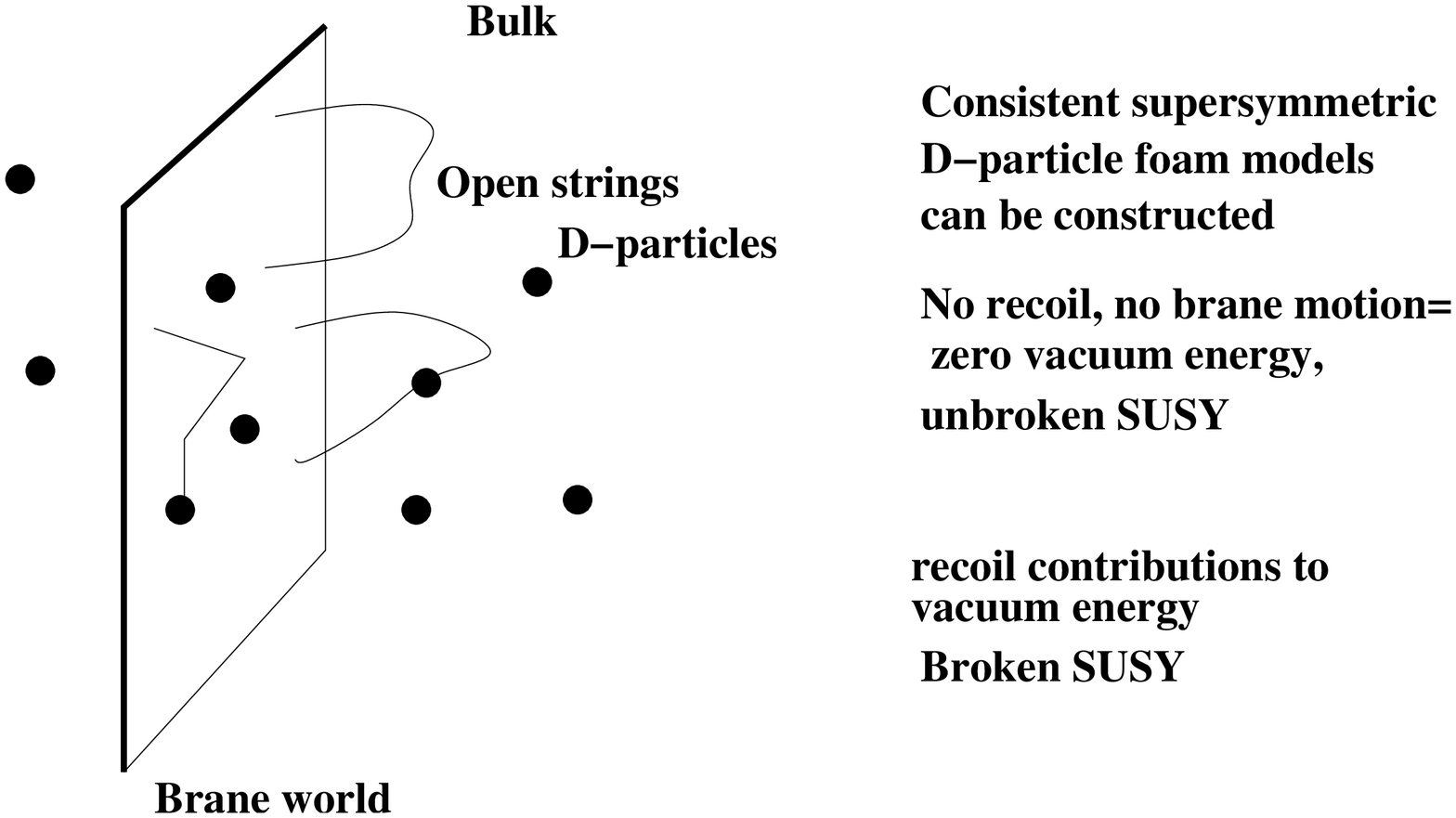} \caption{Schematic
representation of a D-particle space-time foam model. The figure indicates also the capture/recoil
process of a string state by a D-particle defect for closed (left picture) and open
(right picture) string states, in the presence of a D-brane world. The presence of a
D-brane is essential due to gauge flux conservation, since an isolated
D-particle cannot exist. The intermediate composite state at $t=0$, which has
a life time within the stringy uncertainty time interval $\delta t$, of the
order of the string length, is responsible for the distortion of the surrounding
space time during the scattering. This results to \emph{non-trivial optical properties} (refractive index but \emph{not} birefringence) for this space time.}%
\label{fig:recoil}%
\end{figure}
 According to the D-foam model~\cite{Dfoam}, our Universe, after perhaps appropriate compactification, is represented  as a three brane (D3), propagating in a bulk space-time punctured by D0-brane (D-particle) defects. As the D3-brane world moves in the bulk, the D-particles cross it, and for an observer on the D3-brane the situation looks like a ``space-time foam'' with the defects ``flashing'' on and off (``D-particle foam''). The open strings, with their ends attached on the brane, which represent matter in this scenario, can interact with the D-particles on the D3-brane universe in a topologically non-trivial manner, involving splitting and capture of the strings by the D0-brane defects, and subsequent re-emission of the open string state (see fig.~\ref{fig:recoil}).
D-foam can be considered either in the context of type-IIA string theory~\cite{emnnewuncert}, which contains point-like D0-branes, or in the context of the phenomenologically more realistic type-IIB strings~\cite{li}.
However, in the latter case, since the theory admits no D0-branes, one should view the ``D-particles'' as D3-branes (which are allowed) wrapped up (i.e. compactified) around spatial three cycles (in the simplest scenario).

It should be emphasized that in the type IIA model, due to electric charge conservation, only \emph{electrically neutral} matter interacts non-trivially with the D-particle ``foam'', which is thus transparent to charged matter~\cite{ems,emnnewuncert}. On the other hand, in the type IIB-stringy D-foam model~\cite{li}, there are non trivial interactions of charged matter strings, representing electrons and other electrically charged excitations, with the wrapped up D3-branes (``D-particles''). However, such interactions are suppressed by appropriate factors, which --- depending on the construction -- can make these interactions several orders of magnitude smaller than the photon-D-particle interactions. In this way, the anomalous effects of the D-foam on electron dispersion relations can become compatible with the stringent constraints imposed by synchrotron radiation measurements from the Crab Nebula~\cite{crab}.

We therefore observe that the D-brane models of space-time foam do lead to a sort of \emph{violation of the equivalence principle}~\cite{ems}, in the sense that the quantum-gravitational interactions of matter are \emph{not} of \emph{universal} strength among particle species. This is an important aspect, which seems to differentiate the D-foam model from other generic models of quantum gravity, including Deformed Special Relativities, which treat the gravitational interactions of all matter species as universal.
As we shall discuss in the next section, this leads to important phenomenological properties of the D-foam model,
allowing it to evade the stringent astrophysical constraints on modified dispersions linearly suppressed by the quantum gravity scale.

\section{Brane-Foam and High-Energy $\gamma$-Ray Astrophysics}

The interaction and capture of string matter by the D-particle defects can be calculated quantitatively in the first-quantized-string framework by calculating the corresponding string amplitudes of a split matter string, represented in terms of open strings, in the background of a D-particle, that is by considering the open string interactions in the presence of Dirichlet boundary conditions on the world-sheet~\cite{emnnewuncert}. The capture time $\delta t$ amounts to a \emph{causal} time delay due to the formation of an intermediate string state~\cite{sussk1}, which oscillates from zero length to a maximum length, that can be estimated actually by the string uncertainty principles. The fact that the interactions are causal is an exclusive feature of string theory, as emphasized in \cite{sussk1}, and stems from the fact that the amplitude calculations lead to the existence of only retarded waves, emitted after the scattering of the open string states with the defect, and the absence of advanced waves that could cause violations of causality, which by the way do characterize local field theories in non-commutative space times.

For our purposes the existence of causal delays would imply that such delays are additive during the path of a photon from its emission from a distant astrophysical source to the observation point.
During each interaction of a photon, represented as an open string state, with a ``D-particle'' defect in the foam, the string amplitude calculation or the application of the uncertainty principles lead to the estimate for the delayed re-emission after the scattering:
\begin{equation}
\delta t_0 \sim \sqrt{\alpha '} \, E
\label{singledelay}
\end{equation}
where $\alpha ' $ is the Regge slope of the string, related to the fundamental string length ($\ell_s$) and mass ($M_s$) scale by $\sqrt{\alpha '} = \ell_s =1/M_s$, and $E$ is the total energy of the incident string state.
The details of the re-emission have it that there is a series of outgoing waves, which however have attenuating amplitudes, such that practically only the first emitted wave is significant, which leads to the estimate (\ref{singledelay}) for the incurred time delay per interaction of the photon with a ``D-particle'' in  the foam.
The recoil of the defect leads to sub-leading effects, and this is why we ignore it in (\ref{singledelay}). Indeed, as discussed in detail in \cite{emnnewuncert,li}, the recoil induces a non-trivial metric distortion, which is actually formally equivalent to the metric felt by an open string in an electric field background. In the recoil case, the r\^ole of the ``electric'' field is played by the recoil velocity of the D-particle $u_i = \frac{g_s\,\Delta k_i}{M_s} $, $i$ a spatial index, with $\Delta k_i$ the spatial momentum transfer of the string state during its interaction with the D-particle defect of mass $M_s/g_s$. The presence of the recoil leads~\cite{sussk1,emnnewuncert,li} to a modification of (\ref{singledelay}) by: $\delta t_0 \sim \sqrt{\alpha '} \frac{E}{1 - |u_i^2|} $, and therefore the recoil effects are sub-leading, being suppressed by higher powers of the string scale $M_s$~\footnote{A word of caution is in order at this point, concerning the relativistic limit of the recoil velocities $u_i \to 1$ (in units of $c=1$). As becomes clear from the above discussion, for high string mass scales $M_s$ of order $10^{18}$ GeV, which is the traditional string case, the velocity $u_i \ll 1$ for all photon energies currently observed, which are up to order $10^{20}$ eV. This is still eight orders of magnitude smaller than the Planck scale. However, in the modern approach to strings, $M_s$ can be as low as a few TeV, which prompted the enormous recent interest on the possibility of discovering stringy quantum gravity effects at energies reachable at the Large Hadron Collider (LHC). In our context of D-foam, this may open up the possibility of much larger effects coming from the interaction of stringy matter with the massive defects. However, for string theories with such low $M_s$, the correct phenomenology would imply string couplings much smaller than one, such that the combination $M_s/g_s$ is still of order of the four-dimensional Planck scale, $M_P \sim 10^{19}$ GeV. Thus, even in this case, the D-particle recoil velocities, whose magnitude is suppressed by this combination, would be much smaller than one, and their effects can be ignored.}.

The expansion of the Universe affects the magnitude of the observed delay, which differs from (\ref{singledelay}) by terms depending on the red-shift $z$ and the Hubble expansion rate $H(z)$. Indeed, the observed delay from a single scattering event, is affected by two facts: (i) the time dilation factor~\cite{JP} $(1 + z)$ and (ii) the cosmic red-shift~\cite{naturelater}, according to which the observed energy of the photon is $E_{\rm obs} = E_0/(1 + z)$. Thus the observed delay associated with (\ref{singledelay}) is given by:
\begin{equation}\label{obsdelay}
\delta t_{\rm obs} = (1 + z) \delta t_0 = (1 + z)^2 \sqrt{\alpha '} E_{\rm obs}
\end{equation}
If there is a linear density of D-particles $n(z)$, at redshift $z$, then we have $n(z) d\ell =  n(z) dt$ defects per co-moving length (in units of the speed of light \emph{in vacuo} of Special Relativity, $c=1$). The quantity $dt$ denotes the Robertson-Walker infinitesimal time interval of a co-moving observer. Since from  each scattering event of the string matter with a D-particle there is a \emph{causal} delay $\delta t_0$, the total delay of an energetic photon in the region is given by $n(z) (1 + z)^2 \sqrt{\alpha '} E_{\rm obs} \, dt $. One may relate $dt$ to the Hubble rate $H(z)$ in the standard way, $dt = - \, \frac{dz}{(1 + z)\, H(z)}$. The \emph{total observed delay} of a photon from the source at redshift $z$ till observation ($z=0$) is given by the sum of all these infinitesimal delays, given that \emph{each} infinitesimal \emph{delay} due to the individual scattering of a photon with a D-particle defect, encountered in its path, is a \emph{causal event}~\cite{emnnewuncert,naturelater}:
\begin{equation}
\label{totaldelay}
\Delta t_{\rm obs} = \int_0^z dz \frac{n(z)\, E_{\rm obs}}{M_s \,H_0} \,\frac{(1 + z)}{\sqrt{\Omega_M \, (1 + z)^3 + \Omega_\Lambda}}~,
\end{equation}
where we used Standard Cosmology, with $H_0^{-1}$ the present-epoch Hubble time.

It is important to stress that the delays (\ref{singledelay}) are consistent, and in fact \emph{saturate}~\cite{sussk1,emnnewuncert,li}, the space-time stringy quantum uncertainties~\cite{yoneya} $\Delta t \, \Delta X \ge \alpha '$, while a non-zero D-particle recoil velocity $u^i \ne 0$ leads to a space-time non-commutativity~\cite{mavro, li} $[ X^i, t ] \sim \frac{u^i}{1 - |u^i|^2} $.
The presence of statistically significant populations of D-particles in the D-foam, encountered by the photon
 as it traverses large cosmic distances, enhances such uncertainties (c.f. (\ref{totaldelay})) to a point of lying within the sensitivity of current observations, and thus measurable in principle. In this sense, the cosmic sources (AGN, GRB etc.) play the r\^ole of ``Heisenberg microscopes'' for the stringy space-time~\footnote{In generic quantum gravity models, such uncertainties among coordinates alone have also been argued to exist, due to diffeomorphism invariance considerations~\cite{roberts}. Our astrophysical tests described here can also be applied to these more general models.}.

The important difference of the D-foam from other generic and rather phenomenological approaches to quantum gravity, including deformed special relativities, lies on the fact that the total delay (\ref{totaldelay}) is proportional to the linear density of D-particles $n(z)$, which results in an ``effective'' \emph{red-shift dependent} quantum-gravity scale:
\begin{equation}
M_{\rm eff, QG} \equiv \frac{M_s}{n(z)}~.
\label{effqgscale}
\end{equation}
The function $n(z)$ is essentially arbitrary in the context of the D-foam models, as it depends on microscopic details of the higher-dimensional bulk physics.
In certain models, which are perfectly consistent from the point of view of backgrounds of string/brane theory~\cite{Dfoam}, the moving D3-brane world in fig.~\ref{fig:recoil} may encounter an \emph{inhomogeneous} population of bulk D-particle defects. A relative motion between a bulk D-particle and a D3-brane will cause repulsive potentials proportional~\cite{Dfoam} to the relative velocity $v^2$. At late epochs of our Universe, we may imagine a situation in which the D3-brane moves extremely slowly in the bulk, in such a way that no appreciable contributions to the brane vacuum energy arise from such interactions.

It is thus possible to encounter situations, consistent with late-era Cosmological observations, in which, for earlier epochs of the Universe, say $z \ge 1$, the density of D-particles is smaller than that at later eras, $ z < 1$.
Due to the extremely slow brane motion in the bulk, of course, in such situations, the density of D-particles drops by several orders of magnitude in relatively very small bulk regions.
In such a case, then, one can explain -- in the context of D-foam models -- the most recent data on a delayed arrival of energetic photons from the Gamma-Ray Burst 090510, observed by the FERMI Telescope~\cite{grb090510}, with an effective  quantum gravity scale (\ref{effqgscale}), $M_{\rm eff, QG} \gg M_P$, consistently with the MAGIC (and possibly HESS) observations~\cite{MAGIC2,emnnewuncert}, corresponding to an $M_{\rm eff, QG} \sim M_s \sim 10^{18}$~GeV. All one needs is a drop of the effective density of D-particles $n(z)$ by two orders of magnitude as one goes from red-shifts $z=0.03$ (MAGIC) to $z \sim 1$ and higher (FERMI)~\footnote{In such scenarios, for the
case of GRB 080916c, at red-shift $z = 4.2 \pm 0.3$~\cite{grbglast}, only part of the 16 second observed delay of the energetic 13 GeV photons can be attributed to the D-foam, but this is fine.}.

Certainly, many more observations are needed, at various red-shifts, so as to determine, in the context of D-foam models, a suitable function, $n(z)$, for the density of D-particle defects, that could disentangle a possible genuine quantum-gravity effect from source effects that can also cause delayed emissions of the more energetic photons. However, from the above consideration, we hope we made it clear to the reader that linearly-suppressed anomalous dispersion effects due to (stringy) quantum gravity, can certainly survive the recent astrophysical observations, under certain conditions outlined above. From the theoretical side, what one needs is to develop detailed cosmological models of D-foam and compare them with other cosmological observations, such as CMB and others. This is left for the future.

\section{Brane-Foam Effects in Particle Interferometers }

If at late eras of the Universe, one encounters a linear density of D-particle defects of order one defect per string length, with $M_s \sim 10^{18}$~GeV, then such effects can have consequences that are in principle falsifiable in the next upgrades of neutral meson factories, such as an upgrade of DA$\Phi$NE in Frascati Laboratories (Italy)~\cite{adidomenico}. In fact, they constitute rather \emph{unique} signatures of the D-particle foam, pertaining to a specific type of violation of CPT symmetry advocated in \cite{bernabeu}.

These effects are associated with induced modifications of the
Einstein-Podolsky-Rosen (EPR) correlations of entangled states
of neutral mesons in meson factories ($\phi-$ or B-factories)~\cite{bernabeu}.
These modifications concern the nature of the products of the decay
of the neutral mesons in a factory on the two sides of the detector.
For instance, for neutral kaons, if the CPT operator is a well defined
operator, even if it does not commute with the Hamiltonian of the system,
the products of the decay contain states $K_LK_S$ only.
On the contrary, in the case of CPT breakdown
through decoherence (ill defined CPT operator), one obtains
in the final state,
{\it in addition} to $K_LK_S$,
also
$K_SK_S$ and/or $K_LK_L$ states. The strength of such effects, termed the $\omega$-effect~\cite{bernabeu}, has been calculated in \cite{bernsarkar}, within the above-described framework of D-foam models, and found that in some models, the effects can be falsifiable in a possible next upgrade of meson factories, such as DA$\Phi$NE-2~\cite{adidomenico}. In the parametrization of \cite{bernabeu},
such a strength is determined by the magnitude of a complex parameter, $\omega$, which in the model of D-foam, with one D-particle per string length, encountered in the path of a neutral meson, is estimated to be of order~\cite{bernsarkar}:
\begin{equation}\label{finalomega}
 |\omega|^2  \sim ~\frac{g_s^2}{M_s^2}\frac{\left(m_1^2 +  m_2^2\right)}{|m_1 - m_2|^2}p^2~,
  \end{equation}
where $p$ is an average momentum of the Kaon, $M_s=1/\ell_s$ is the string mass scale, and, as appropriate for Da$\Phi$NE $\phi$-factory~\cite{adidomenico}, we assumed non relativistic dispersion relations for the neutral kaons, of masses $m_1$ and $m_2$.

Since, for neutral Kaons in a $\phi$ factory, we have
$m_1 - m_2 = m_{L} - m_{S} = \sim3.48 \times10^{-15}~\mathrm{GeV}~,$
and the momenta are of order $1$ GeV,
the result (\ref{finalomega}) implies the estimate
\begin{equation}\label{dfoamomega}
|\omega|_{\rm D-foam} = {\cal O}\left( 10^{-5}\right)~,
\end{equation}
for string mass scales $M_s \sim  10^{18}$~GeV, as required by the D-foam interpretation of the MAGIC astrophysical observations on delayed arrival of energetic photons~\cite{MAGIC2}, and string couplings $g_s$ such that $M_s/g_s \sim M_P \sim 10^{19}$~GeV.

At present, the estimate (\ref{dfoamomega}) is still some two
orders of magnitude away from the current bounds of the $\omega$-effect
by the KLOE collaboration at DA$\Phi$NE~\cite{adidomenico}, giving $|\omega| < 10^{-3}$,
but it is within the projected sensitivity of the proposed upgrades.
The above estimate is valid for non-relativistic meson states, where each meson has been treated as a \emph{structureless} entity when considering its interaction with the D-foam.
The inclusion of details of the Kaons strongly-interacting substructure may, of course, affect the results significantly. Nevertheless, the above example indicates how two entirely different physical tests, pertaining to diverse physics (astro and particle physics) can provide complementary information to each other, regarding this particular model of stringy space-time foam.

\section{Conclusions}

The point of the talk was to argue that the field of phenomenology of quantum gravity can become realistic, and already many existing models can be falsified by a plehtora of diverse experiments, covering entirely different sectors of modern physics: from observations of high-energy $\gamma$ rays from cosmic sources to particle interferometers.
The field of Gamma-Ray astrophysics, in particular after the operation or launch of precision Telescopes or arrays of them, have recently provided us with some spectacular events, with a delayed arrival of more energetic photons from a source, which naively appear to reach or exceed the sensitivity to the so-called Planck length (or mass) scale.
In this sense, it has been claimed that models of quantum gravity with linear suppression of anomalous dispersion effects have already been excluded by the recent GRB 090510, detected by the Fermi Telescope in May 2009.
We have argued in this presentation, that this is not so, at least at present.

In the context of a specific stringy model of space-time foam, we have supported the point of view that high-energy $\gamma$-ray observations from distant cosmic sources can provide us with ``Heiseinberg-like Microscopes'', which can test space-time uncertainties that could be characteristic of quantum (and possibly stringy) space-time effects.
In such models there is an effective quantum gravity (mass) scale, which is not simply given by a constant Planck or string mass scale, but rather by a red-shift dependent quantity, in general, (\ref{effqgscale}), which is inversely proportional to the density of ``foamy'' structures in our Universe at a given epoch. In the particular model of ``D-foam'', which was the basis of our discussion, such structures are provided by point-like (true or ``compactified'', depending on the model) D-branes that populate the late eras of our Universe.

Certainly, at present, the existence of a  scarce set of high-energy cosmic photon data, available to date, \emph{prevents} one from drawing safe conclusions or disentangling potentially true quantum-gravity effects from source effects that could also be responsible for the observed delays of the more energetic photons. Nevertheless, what we would like to demonstrate with this talk is that the era of quantum gravity phenomenology has already reached us, and the future appears to be bright and exciting.

\section*{Acknowledegments}
I would like to thank the organisers of this stimulating meeting (XXV Max Born Symposium, Wroclaw (Poland), 29th June- 3rd July 2009) for the invitation.
This work is partially supported by the European Union
through the Marie Curie Research and Training Network \emph{UniverseNet}
(MRTN-2006-035863).

\end{document}